\documentstyle[prl,twocolumn,aps,floats,epsfig,amsmath,amssymb]{revtex}
\begin{document}

\twocolumn[
\hsize\textwidth\columnwidth\hsize\csname@twocolumnfalse\endcsname
\draft

\title{Low density spin-polarized transport in 2D semiconductor
  structures: The enigma of temperature dependent magnetoresistance of
  Si MOSFETs in an in-plane applied magnetic field}
\author{S. Das Sarma and E. H. Hwang}
\address{Condensed Matter Theory Center, 
Department of Physics, University of Maryland, College Park,
Maryland  20742-4111 } 
\date{\today}
\maketitle

\begin{abstract}
The temperature dependence of 2D magnetoresistance in an applied
in-plane magnetic field is theoretically considered for electrons in
Si MOSFETs within the screening theory for long-range charged impurity
scattering limited carrier transport. In agreement with recent
experimental observations we find an essentially temperature
independent magnetoresistivity for carrier densities well into the 2D
metallic regime due to the field-induced lifting of spin and,
perhaps, valley degeneracies. 
In particular the metallic temperature dependence of the ballistic
magnetoresistance is strongly suppressed around the zero-temperature
critical magnetic field ($B_s$) for full spin-polarization, with the
metallic temperature dependence strongest at $B=0$, weakest around $B
\sim B_s$, and intermediate at $B \gg B_s$.

\vspace*{.5cm}
\end{abstract}
]

\section{introduction}

The phenomena of 2D metallic behavior and the associated 2D
metal-insulator transition (MIT) comprise a complex set of
low-temperature transport behavior of low-density 2D electron (or
hole) systems in high-quality (i.e. low-disorder and high-mobility)
semiconductor structures. In particular, the 2D resistivity,
$\rho(n,T,B)$, shows intriguing and anomalous dependence on carrier
density ($n$), temperature ($T$), and applied magnetic field ($B$)
parallel to the 2D plane \cite{review1,review}. 
For example, the low temperature $\rho(T)$,
in zero applied field, shows remarkably strong ``metallic''-like
(i.e. $d\rho/dT >0$ for $n>n_c$) temperature dependence for 2D carrier
densities above the so-called critical density ($n_c$) for the 2D MIT
whereas, for $n<n_c$, the system exhibits insulating behavior
($d\rho/dT <0$). The strength of the metallicity (i.e. how strongly
$\rho(T)$ increases with $T$ at low temperatures for $n>n_c$) and the
actual value of $n_c$ are highly nonuniversal, and depend strongly
(and nontrivially) on the material and on the 2D sample quality.
The application of an in-plane magnetic field $B$ has several
interesting effects on the 2D metallic 
phase \cite{b1,b2,b3,b4,tsui,shashkin,b5,b6}:
(1) at a fixed low $T$, 
the system develops a large positive magnetoresistance with $\rho(B)$
increasing very strongly (by as much as a factor of 4) with $B$ upto a
maximum field $B_s$; (2) for $B>B_s$, $\rho(B)$ either saturates (or
increases slowly with $B$ for $B>B_s$) showing a distinct kink at
$B=B_s$; (3) the temperature dependence of resistivity is considerably
suppressed in a finite parallel field; (4) for 2D electrons in Si
MOSFETs, which is the most extensively studied system in the context
of 2D MIT, the external magnetic field at densities close to the
critical density (i.e. $n \ge n_c$) seems to drive the system into a
strongly insulating phase generating a huge positive magnetoresistance
and manifesting the strongly
insulating activated temperature dependence typical of the zero-field
transport for $n<n_c$.

In this paper we restrict ourselves to the highly metallic
(``ballistic'') $n>2n_c$ regime and investigate theoretically the
enigmatic temperature dependence of the parallel field
magnetoresistatce $\rho(B;T)$ of n-Si MOSFETs as described in items
(1) --(3) above. The item (4), which arises from the destruction of
the 2D effective metallic phase by a large in-plane field in the
$n\ge n_c$ regime, is obviously related to the 2D metal-insulator
transition itself (or more precisely, its parallel field dependence)
and is beyond the scope of the current theory which deals only with the
apparent effective 2D metallic phase in the ballistic regime. We
include in the theory only resistive carrier scattering by screened
effective disorder arising from charged impurity centers randomly
distributed at the Si-SiO$_2$ interface with the screening calculated
within the finite wave vector self-consistent field
random-phase-approximation (RPA) 
at finite temperatures fully incorporating
the carrier spin polarization induced by the applied in-plane
magnetic field. In our theory the magnetic field effects enter only
through the carrier spin polarization correction. 
The temperature dependence (at zero field) \cite{DH2004} and the
parallel-field  dependence (at $T=0$) \cite{gold_b} of $\rho(T;B)$
have earlier been individually theoretically studied in the literature
for Si MOSFETs within the screening theory formalism -- our goal here
is to combine the two to obtain a full theory for $\rho(T;B)$ in Si
MOSFETs including spin-polarization effects on the finite-temperature
screening properties.

We emphasize that, although the 2D ``metallic'' transport properties
are anomalous in the presence of a parallel magnetic field in all of
the 2D systems studied so far, the 2D Si MOS system is unique in
exhibiting a particularly intriguing temperature dependent
magnetoresistance $\rho(T,B)$ where $\rho(T)$ for $B>B_s$ and $n \ge
1.5 n_c$ seems to be essentially a constant at low temperatures
\cite{tsui,shashkin}. In 
this paper we propose a simple physically motivated theoretical
explanation for the anomalous transport properties of Si MOSFETs in
the presence of a strong parallel magnetic field. Our explanation is
based on the substantial modification of the effective quenched
disorder in the system (arising from the screened background impurity
potential) due to the applied parallel field which strongly affects
the 2D screening properties.
We believe (and show in this paper) that a qualitative understanding
of the temperature dependence of $\rho(T;B)$ can be obtained on the
basis of an effective field-dependent disorder.

The applied parallel magnetic field has, in principle,  seven distinct
(and sometimes, opposing) physical effects on the 2D metallic
transport behavior: (1) the magneto-orbital coupling \cite{DH_B} 
due to the finite thickness of the 2D layer leads to an
anisotropically increasing 2D effective mass, consequently giving rise
to a positive anisotropic magnetoresistance; (2) the parallel field
couples the 2D in-plane dynamics to the dynamics in the confinement
direction (perpendicular to the 2D plane) leading to intersubband
scattering (i.e. 2D to 3D crossover) induced positive
magnetoresistance \cite{DH_B}; 
(3) the parallel field may effectively ``enhance''
weak localization ``insulating'' temperature correction \cite{Lewalle}
by suppressing
various ``metallic'' contributions (e.g. screening) to the temperature
dependent resistivity; (4) the parallel field-induced carrier spin
polarization leads to an enhancement of the effective Fermi momentum,
$k_F(B)$, which tends to contribute a {\it negative} magnetoresistance
through the decrease of the charged impurity scattering matrix
elements; (5) the parallel field-induced carrier spin polarization
reduces the strength of screening (by as much as a factor of 2) since
the electronic density of states decreases from 2 to 1 as the 2D
electrons become completely spin-polarized; (6) the parallel field may
further suppress screening (by as much as an additional factor of 2) in
Si MOSFETs by lifting the valley degeneracy factor (for example, from
2 to 1 if $\Delta_v >E_F$, $k_BT$, where $\Delta_v$ is the valley
splitting in the presence of the parallel field);
(7) the parallel field induced modification of the electron-electron
interaction \cite{ZNA}, due to the spin-polarization of the 2D
electrons which suppresses the ``triplet'' interaction channel.

We develop a theory  for $\rho(T,B,n)$ in Si MOSFETs
including the three physical mechanisms [(4) -- (6) listed above]
which are important for Si MOS systems. We leave out the
magneto-orbital effects (items (1) and (2) above) \cite{DH_B} 
because the
magneto-orbital corrections are rather small in Si MOSFETs since the
quasi-2D layer width (i.e. the confinement size transverse to the 2D
plane) is rather small, making magneto-orbital coupling essentially
negligible in Si MOSFETs except at very high parallel fields. We do
not consider the weak localization correction since it is
straightforward to include it (in an ad hoc fashion) if experimental
results warrant such a theoretical adjustment \cite{Lewalle}.
We also uncritically ignore all electron-electron interaction
corrections beyond the screening (i.e. ring diagrams) effects, arguing
screening to be the dominant physical mechanism controlling transport
limited by Coulomb disorder arising from charged impurity scattering.
Interaction effects in the presence of a parallel magnetic field have
been considered in the literature \cite{ZNA}.

The main physical effect we consider, namely the parallel field
induced decrease of the 2D density of states leading to a strong field
induced suppression of screening (and hence a strong positive
magnetoresistance), has earlier been discussed in the literature
\cite{gold_b}, primarily in the context of the $T=0$ magnetoresistance
itself (whereas our focus is on the temperature dependence of
magnetoresistance). We have recently \cite{DH_S} discussed the
similarity between the behavior of $\rho(B;T=0)$ and $\rho(T;B=0)$ in
Si MOSFETs \cite{pudalov01} as arising from the field-induced or the
temperature-induced suppression of screening, leading to qualitatively
``similar'' parallel-field dependent (at $T=0$) and
temperature-dependent (at $B=0$) effective disorder in MOSFETs. (As an
aside we mention that the situation is quite different in 2D GaAs
electron \cite{zhu} and hole \cite{noh} systems where substantial
magneto-orbital coupling is present, leading to qualitatively
different $\rho(B)$ and $\rho(T)$ behaviors.) The goal of the current
paper is to develop a theory for $\rho(T;B)$, the {\it temperature
  dependence} of the 2D magnetoresistivity in Si MOSFETs, particularly
at large fields $B\ge B_s$, where the 2D system is presumably
completely spin-polarized. The motivation for our theoretical work
comes from the recent experimental work \cite{tsui}
reporting a puzzling absence
of any temperature dependence in $\rho(T;B>B_s)$ of Si MOSFETs in the
completely spin-polarized large applied parallel field ($B > B_s$)
ballistic ($n > 1.5n_c$) transport regime. Since the zero-field (i.e.,
spin unpolarized) 2D transport in these high-quality low-density Si
MOSFETs is characterized by strong metallicity, i.e. a strong metallic
temperature dependence in $\rho(T)$, the apparent complete suppression
of the metallicity in the spin-polarized system is intriguing and has
attracted a great deal of attention. The complete reported absence of
any temperature dependence of $\rho(T;B>B_s)$ in Si MOSFETs
\cite{tsui} also
stands in sharp contrast to the corresponding situation \cite{sige} in
n-Si/SiGe 2D structures, where $\rho(T;B>B_s)$ shows metallic
temperature dependence in the spin-polarized case, albeit with a
reduced magnitude of $d\rho/dT$ in accordance with the screening theory.
We discuss these issues in great details in the next four sections of
this article. In sec. II we provide
the Boltzmann transport theory. In sec. III we give our calculated
results. In sec. IV
we discuss our results. We conclude our paper in sec. V.

\section{Theory}

We use the semiclassical Boltzmann transport 
theory \cite{DH2004} including only the
effect of resistive scattering by random charged impurities at the
Si-SiO$_2$ interface -- at low carrier densities (and at low
temperatures where phonon scattering is unimportant) of interest in
the 2D MIT problem, transport in high-mobility Si MOSFETs is known to
be predominantly (but perhaps not entirely) limited by long-range
charged impurity scattering. The density 
of the random interface
charged impurity centers is therefore the only unknown parameter in
our model, which sets the overall scale of $\rho(T=0)$ without 
affecting the $(T,n,B)$ dependence of resistivity that is of
interest in the problem. The bare disorder potential arising from
oxide charged impurity centers being long-range Coulombic in nature,
the most important physics ingredient (at the zeroth-order) that any
transport theory must incorporate is the regularization of the {\it
  bare} long-range Coulombic disorder potential by screening the
impurity potential. Within our zeroth-order minimal transport theory,
this is precisely what we do by calculating the finite temperature
screened effective disorder through the static (finite-temperature and
finite-wave vector) RPA screening of the bare long-range Coulombic
disorder \cite{DH2004}. 
We then use the Boltzmann theory to calculate the
finite-temperature carrier resistivity limited by scattering due to
the effective ``regularized'' (i.e. screened) impurity disorder. We
include in our theory the realistic effects of the {\it quasi}-2D
layer width of Si MOSFETs through the appropriate quantum form
factors, and most importantly we include the nonperturbative effect of
the parallel magnetic field through the spin polarization of the 2D
electron system fully incorporating the physical mechanisms (4) and
(5) discussed above \cite{gold_b,DH_S} in Sec. I. 
(The inclusion of the mechanism (6) in the
theory, namely the field-induced lifting of valley degeneracy,
i.e. valley polarization, is straightforward by adjusting the valley
degeneracy factor $g_{\nu}$ in the 2D density of states and is
discussed below.)

The formal aspects of the calculation for $\rho(n;T;B)$ in the
presence of the parallel magnetic field are
essentially the same as those for $\rho(n;T)$ at $B=0$ except for the
existence of different spin-polarized carrier densities $n_{\pm}$.
When the parallel magnetic field is applied to the system the
carrier densities $n_{\pm}$ for spin up/down are not equal. Note that 
the total density $n=n_+ + n_-$ is fixed. 
The spin-polarized densities
themselves are obtained from the relative shifts (i.e. the
spin-splitting) in the spin up and down bands introduced by the Zeeman
splitting associated with the external applied field $B$. At $T=0$,
one simply has $n_{\pm} = n(1\pm B/B_s)/2$ for $B \le B_s$ with $n_+ =
n$ and $n_-=0$ for $B\ge B_s$ (and $n_+=n_-=n/2$ at $B=0$), where
$B_s$, the so-called saturation (or the spin-polarization) field, is
given by $g\mu_B B_s = 2E_F$ where $g$ is the electron spin $g$-factor
and $\mu_B$ the Bohr magneton. For $T\neq 0$, $n_{\pm}$ is determined
using the Fermi distribution function in the standard manner.
Thus, in the presence of the magnetic field
the total conductivity can be expressed as a sum of spin
up/down contributions
\begin{equation}
\sigma =\sigma_+ + \sigma_-,
\end{equation}
where $\sigma_{\pm}$ is the conductivity of the ($\pm$) spin subband.
The total carrier resistivity $\rho$ is defined by $\rho \equiv
1/\sigma$. The conductivities $\sigma_{\pm}$ are given by
\begin{equation}
\sigma_{\pm} = \frac{n_{\pm}e^2 \langle \tau_{\pm}\rangle }{m},
\end{equation}
where $m$ is the carrier effective mass, and the energy averaged
transport relaxation time $\langle \tau_{\pm} \rangle$ for the ($\pm$)
subbands are given in the Boltzmann theory by
\begin{equation}
\langle \tau_{\pm} \rangle  = \frac{\int d\varepsilon
  \tau(\varepsilon) \varepsilon \left [ -\frac{\partial
  f^{\pm}(\varepsilon)}{\partial \varepsilon}  \right ]} 
  {\int d\varepsilon \varepsilon \left [ 
    -\frac{\partial f^{\pm}(\varepsilon)}{\partial \varepsilon} \right
  ]},
\label{tau}
\end{equation}
where $\tau(\varepsilon)$ is the energy dependent relaxation
time, and $f^{\pm}(\varepsilon)$ is 
the carrier (Fermi) distribution function 
\begin{equation}
f^{\pm}(\varepsilon) =\frac{1}
{1+e^{\beta[\varepsilon-\mu_{\pm}(T)]}},
\end{equation}
where $\beta \equiv (k_BT)^{-1}$, and $\mu_{\pm} = \frac{1}{\beta} \ln
\left [ -1 
  + \exp(\beta E_{F^{\pm}}) \right ]$
is the chemical potential for the up/down spin state (with Fermi energy
$E_{F^{\pm}}$) at finite temperature.
The spin-polarized transport can then be calculated within this
two-component (spin-up and -down) carrier system (without any
spin-flip scattering since the bare impurities are non-magnetic), with
the screening of the bare disorder being provided by both spin up and
down carriers.

Within the framework of linear transport theory the relaxation time
for a carrier with energy $\varepsilon_k$ is given by
\begin{eqnarray}
\frac{1}{\tau(\varepsilon_k)} = \frac{2\pi}{\hbar}\sum_{\alpha,{\bf k}'}
\int^{\infty}_{-\infty}dz N_i^{(\alpha)}(z)
|u^{(\alpha)}({\bf k}-{\bf k}';z)|^2 \nonumber \\
\times (1-\cos \theta_{{\bf k k}'})
\delta(\varepsilon_k-\varepsilon_{k'}), 
\label{itau}
\end{eqnarray}
where $\theta_{{\bf k k}'}$ is the scattering angle between
wave vectors {\bf k} and ${\bf k}'$ and $\varepsilon_k = \hbar^2 k^2/2m$;
the screened scattering potential is denoted by $u^{(\alpha)}({\bf
  q};z)$ with ${\bf q} \equiv
{\bf k} - {\bf k}'$ and $z$ is 
the confinement direction normal to the 2D layer. 
$N_i^{(\alpha)}(z)$ in Eq. (\ref{itau}) is the 3D charged
impurity density of the $\alpha$-th kind of charged center.
Here we have assumed that the charged centers are distributed
completely at random in the Si-SiO$_2$ interface for MOSFETs.
The screened impurity potential $u^{(\alpha)}({\bf q};z)$ is given by: 
\begin{equation}
u^{(\alpha)}(q;z) =  \frac{2\pi Z^{(\alpha)}e^2}{\bar \kappa q
  \epsilon(q)} F^{(\alpha)}_{\rm i}(q;z), 
\end{equation}
where $Z^{(\alpha)}$ is the impurity charge strength, $\bar\kappa$ is the
background (static) lattice dielectric constant, and $F_{\rm i}$ is a
form factor determined by the location of the impurity and the subband
wavefunction $\psi(z)$ defining the 2D confinement. 
The finite wave vector dielectric screening function $\epsilon(q)$ 
is written in the RPA as 
\begin{equation}
\epsilon(q) = 1- \frac{2\pi e^2}{\kappa q} f(q) \Pi(q,T),
\end{equation}
where $f(q)$ is the Coulomb form factor arising from the subband
wavefunctions $\psi(z)$. In the strict 2D limit $f(q) = 1$.
$\Pi(q,T)$ is 
the 2D irreducible finite-temperature (and finite wave vector)
polarizability function and is give by
$\Pi(q;T)=\Pi_+(q;T) + \Pi_-(q;T)$, where
$\Pi_{\pm}(q;T)$ are the 
polarizabilities of the polarized up/down spin states ($\pm$).
At finite temperature we have
\begin{equation}
\Pi_{\pm}(q,T) = \frac{\beta}{4}\int^{\infty}_{0}d\mu'
\frac{\Pi_{\pm}^0(q,\mu')}{\cosh^2 \frac{\beta}{2}(\mu_{\pm}-\mu')},
\label{piqt}
\end{equation}
where $\Pi_{\pm}^0(q,E_F^{\pm})
\equiv \Pi_{\pm}^0(q)$ 
is the zero-temperature noninteracting static polarizability,
given by 
\begin{equation}
\Pi_{\pm}^0(q) = N_F \left [ 1- \sqrt{1-\left ( {2k_{F^{\pm}}}/{q}
    \right )^2} \theta(q-2k_{F^{\pm}})\right ],
\end{equation}
where $N_F = g_vm/2\pi$ is the density of states per spin 
at Fermi energy, and
$k_{F^{\pm}} = (4\pi n_{\pm}/g_v)^{1/2}$ is the 2D Fermi wave vector
for the spin up/down carriers.
Note that $g_v$ (=2 in the usual $B=0$ Si MOS case) is the valley
degeneracy, and the spin degeneracy, by definition, is assumed to be
lifted by the in-plane field $B$, the usual unpolarized $B=0$
paramagnetic case being $k_{F^+}=k_{F^-}$; $n_+=n_-=n/2$.

We mention that Eqs. (1) -- (9) for the finite-temperature carrier
transport properties in 2D systems comprise a complex set of
multi-dimensional integrals
along with the determination of the quasi-2D Coulomb form factors as
well as the chemical potential of the system.
In the asymptotic regimes of $T/T_F \ll 1$ and $T/T_F \gg 1$, Eqs. (1)
-- (7) simplify (as discussed in ref. \onlinecite{DH2004}) giving rise
to simple analytic behaviors in $\rho(T) \sim O(T/T_F)$ for $T/T_F \ll
1$ and $\rho(T) \sim O(T_F/T)$ for $T/T_F \gg 1$, but these asymptotic
analytic behaviors are of limited experimental relevance since few
experimental parameter regimes satisfy the required conditions
\cite{DH2004} for the asymptotic behavior. Also, at the lowest
experimental temperatures $\rho(T)$ invariably saturates essentially
in all experiments. For arbitrary $T$ and $n$ (as well as $B$) one
must evaluate Eqs. (1) -- (7)  with sufficient accuracy to obtain
$\rho(T,n,B)$ for 2D systems.

The most salient aspects of the parallel field induced carrier spin
polarization are an enhancement of the 2D Fermi wave vector by a
factor of $\sqrt{2}$ and a suppression of the 2D screening
(``Thomas-Fermi'') wave vector by a factor of 2 as the unpolarized
(``paramagnetic'') 2D system becomes completely spin-polarized
(``ferromagnetic'') with the parallel field increasing from $B=0$ to
$B\ge B_s$ (with $B_s$ being the full or saturation spin polarization
field). In the most naive theoretical level one can write (at $T=0$)
$\rho \propto (q_{TF} + 2k_F)^{-2}$ in the strong screening limit
for scattering by screened charged
impurities, leading to 
\begin{equation}
\frac{\rho(B\ge B_s)}{\rho(0)} \le \left ( \frac{q_{TF}(0) +
    2k_F(0)}{q_{TF}(B_s) + 2k_F(B_s)} \right )^2.
\end{equation}
In the usual range of experimental 2D densities $q_{TF} \gg 2k_F$
(``strong screening'') in Si MOSFETs, and therefore
$\rho(B\ge B_s)/\rho(0) \le [q_{TF}(0)/q_{TF}(B_s)]^2 \le 4$. (As an
aside, we note that in the opposite limit of weak screening, $q_{TF}
\ll 2k_F$, which may be approximately the situation in the
high-density 2D n-GaAs system, one gets $\rho(B_s)/\rho(0) \le 1/2.$)
At finite temperatures and with finite wave vector screening
$\rho(B\ge B_s)/\rho(0)$ is expected to be less than 4 in Si MOSFETs
as observed experimentally.
Note that the 2D strong screening condition, $q_{TF} \gg 2k_F$, occurs
at {\it low} carrier densities since $k_F \propto \sqrt{n}$ and
$q_{TF}$ is independent of density in the lowest order.

The suppression of screening due to the parallel field-induced spin
polarization has direct consequences for the temperature dependence of
the magnetoresistivity $\rho(T,B)$ since the temperature dependence of
screening gives rise to an effective metallic behavior of
$\rho(T)$ at zero magnetic field \cite{DH2004,DH_S}. 
In particular, the strength of the
metallicity, i.e. how strong the low temperature metallic temperature
dependence is in a 2D system, depends on the dimensionless parameter
$(q_{TF}/2k_F)^2 \sim (4.2/\tilde{n}) (g_s g_{\nu})^3$ in Si MOSFETs where
$\tilde{n}$ is the carrier density $n$ measured in the units of
$10^{11}cm^{-2}$ and $g_s$, $g_{\nu}$ are respectively the spin and
valley degeneracy factors (with $g_s=2$ and $g_{\nu}=2$ being the
usual zero-magnetic field spin and valley unpolarized case). As the
parallel field increases, $0 \le B \le B_s$, the 2D carriers become
spin-polarized with $g_s$ decreasing from 2 (for $B=0$) to 1 ($B\ge
B_s$), consequently suppressing the metallicity parameter
$(q_{TF}/2k_F)^2$ by a factor 8 between $B=0$ and $B \ge B_s$. This
implies that the temperature dependence of the finite
field resistivity $\rho(T,B,n)$ 
for $B\ge B_s$ at a particular carrier density $n_B$
will be approximately (and qualitatively) similar to the corresponding
{\it zero-field} metallic temperature dependence at a carrier density
$n_0$ which is roughly 8 times higher: $n_0 \approx 8n_B$. Since the
2D metallic temperature dependence is strongly suppressed by
increasing density, these simple screening considerations immediately
suggest a very strong suppression of 2D ``metallicity'' (i.e. the
temperature dependence of 2D resistivity) at high parallel
fields. An equivalent physical way of describing this strong screening
induced (through the spin-polarization dependence of screening)
suppression of the temperature dependence of 2D parallel field
magnetoresistivity is to observe that the metallicity parameter in the
2D n-GaAs system (where $g_{\nu}=1$) is much smaller, $(q_{TF}/2k_F)^2
\sim (0.9/\tilde{n})g_s^3$, due to the much smaller electronic carrier
effective mass in GaAs ($m_{GaAs} =0.067 m_e$ and $m_{Si} = 0.19 m_e$),
and therefore in the presence of a strong parallel field ($B\ge B_s$ so
that $g_s =1$) the Si MOSFET system at a particular carrier density
$n_{Si}$ has a ``metallicity'' which is roughly equal to the
corresponding {\it zero-field} (i.e. $g_s =2$) metallicity in the 2D
n-GaAS system at a density $n_{GaAs} \approx 2n_{Si}$. Since the
observed metallic behavior (i.e. the temperature dependence of $\rho$)
in 2D n-GaAs system is extremely weak except \cite{lilly} 
at very low electron
densities (below $10^{10}cm^{-2}$), the temperature dependence of the
magnetoresistivity $\rho(T,B)$ in Si MOSFETs would thus be strongly
suppressed at large parallel fields. We emphasize that the same
physical mechanism, namely the suppression of screening due to carrier
spin-polarization, leading to the strong positive magnetoresistance
$\rho(B)$ at a fixed low temperature also leads to the strong
suppression of the temperature dependence of $\rho(B,T)$ at a finite
parallel field.

\begin{figure}
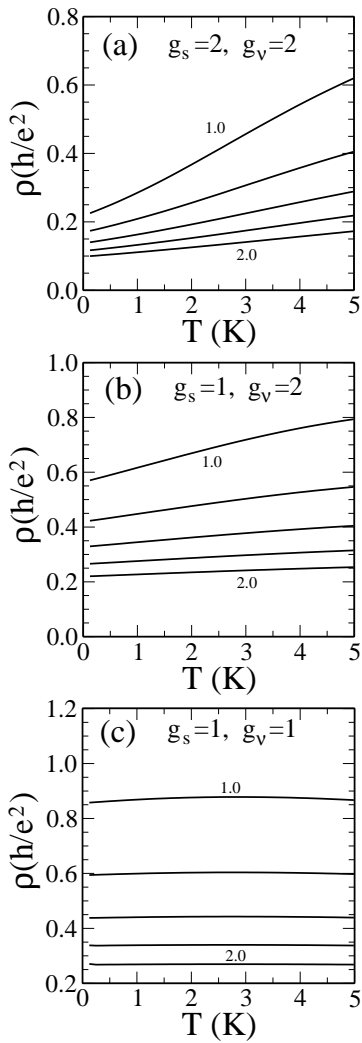

\epsfysize=1.8in
\centerline{\epsffile{fig1a.eps}}
\centerline{\epsffile{fig1b.eps}}
\centerline{\epsffile{fig1c.eps}}
\vspace{0.2cm}
\caption{
The calculated resistivity of Si MOSFET systems as a function of the
temperature for various densities, $n=1.0$, 1.25, 1.5, 1.75,
2.0$\times 10^{11} cm^{-2}$ (top to bottom), (a) for  $g_s=2$,
$g_{\nu}=2$, (b) for $g_s=1$, $g_{\nu}=2$, and (c) for $g_s=1$,
$g_{\nu}=1$.
}
\label{fig1}
\end{figure}


\section{results}

In Fig. 1 we show our calculated temperature dependence of n-Si MOS 2D
resistivity, $\rho(T)$, for several carrier densities in the unpolarized
zero-field case ($g_s=2$, $g_{\nu}=2$), Fig. 1(a) as well as the high
field ($B\gg B_s$) fully spin polarized case ($g_s=1$, $g_{\nu}=2$),
Fig. 1(b). (All magneto-orbital effects \cite{DH_B}
have been ignored in these
calculations.) A comparison of Figs. 1(a) and (b) immediately
demonstrates the strong suppression of 2D metallicity (i.e. the
temperature dependence of $\rho$) in the high-field spin-polarized
system, particularly at higher ($n>10^{11}cm^{-2}$) carrier
densities. At lower 2D densities, however, our calculated $\rho(T)$ in
the spin-polarized system seems to manifest stronger metallicity than
that observed recently by Tsui {\it et al.} \cite{tsui} 
and by Shashkin {\it et al.} \cite{shashkin}
who have  measured the temperature dependence of 2D
magnetoresistivity in low-density Si MOSFETs,
reporting essentially no temperature dependence in the spin polarized
Si MOS systems.

One possible reason for
this weaker experimental temperature dependence of $\rho(T,B)$ could
be that the strong parallel field lifts the valley dependency (i.e.,
$g_{\nu}=1$ in the high field situation) as well as the spin
degeneracy. The long-standing valley degeneracy problem in Si MOSFETs
is poorly understood theoretically except that it is experimentally
well-established that the valley degeneracy is typically lifted by a
small valley splitting, similar to spin splitting,
$\Delta_v$ of poorly understood theoretical
origin. It is also established experimentally that $\Delta_v$
increases with decreasing carrier density and increasing
magnetic field \cite{pudalov_g}. 
We speculate that it is possible that
at low carrier densities in the presence of the parallel magnetic
field the valley degeneracy is lifted ($\Delta_v > E_F$) so that the
Si MOS system becomes a spin and valley polarized system ($g_s =
g_{\nu} =1$) for $B \gg B_s$. This is not unreasonable since the
suppression of screening at large parallel fields would lead to strong
many-body enhancement of valley splitting leading perhaps to the
lifting of valley degeneracy at low densities and high fields.
(Whether the valley degeneracy is indeed lifted in the experimental
MOSFETs in the presence of a strong in-plane field $B\gg B_s$ can only
be ascertained experimentally -- we are here suggesting only the
theoretical possibility.) 
In Fig. 1(c) we show our calculated high field Si MOS $\rho(T)$ in the
spin and valley polarized situation, to be compared with the
spin-polarized (but valley-unpolarized) case in Fig. 1(b) and both
spin and valley unpolarized (i.e. $B=0$) case in Fig. 1(a). The spin
and valley polarized theoretical results in Fig. 1(c) are remarkably
similar to the recent experimental results of Tsui {\it et
  al}.\cite{tsui} -- in 
fact, our results in Fig. 1(c) even reproduce the experimental
observation, as noted in ref. \onlinecite{tsui}, of a weak negative
(i.e. $\rho(T)$ decreasing with increasing $T$) temperature dependence
of $\rho(T)$ arising from the rather small value of $q_{TF}/2k_F$ is
the full spin and valley polarized situation which leads to a
sub-leading negative temperature contribution to $\rho(T)$.

\begin{figure}
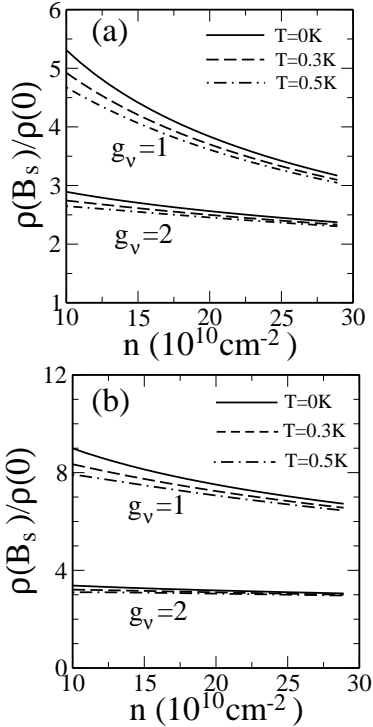

\epsfysize=1.9in
\centerline{\epsffile{fig2a.eps}}
\centerline{\epsffile{fig2b.eps}}

\caption{
Calculated $\rho(B_s)/\rho(0)$ as a function of density (a) for
quasi-2D system and (b) for pure 2D system. Here, at $B=0$, $g_s=g_v=2$.
Note at $B=B_s$ the spins are completely polarized. The lines
corresponding to $g_v=1$ (2) indicate that the valley degeneracy is
(not) lifted at $B=B_s$.
}
\end{figure}

In Fig. 2 we compare $\rho(T)$ at different spin and valley
degeneracies ($g_s=g_{\nu}=2$; $g_s=1$, $g_{\nu}=2$; $g_s =
g_{\nu}=1$) at different densities, with the observation that our
$g_s=g_{\nu}=1$ results are in good agreement with the recent
experimental Si MOS data in the high field ($B > B_s$) situation
\cite{tsui,shashkin}. Our 
speculation of the Si valley degeneracy being lifted ($g_{\nu}=1$), in
addition to the spin degeneracy, for $B >B_s$ could be further tested
by considering $\rho(B)$ at a fixed temperature, as shown in
Fig. 2. At $T=0$ and in the unrealistic and incorrect strictly 2D limit
the lifting of both spin and valley degeneracies will lead to
$\rho(B_s)/\rho(0) \rightarrow 16$ as $n \rightarrow 0$
[Fig. 2(b)]. This will be in direct disagreement with experimental
observations where $\rho(B_s)/\rho(0) < 4$ in Si MOS systems. But, in
the realistic quasi-2D systems (and at finite temperatures) we find
[Fig. 2(a)] that $\rho(B_s)/\rho(0) < 4$ 
for $n \ge 2.0 \times 10^{11} cm^{-2}$
-- for $n < 2.0 \times 10^{11} cm^{-2}$, 
we find $\rho(B_s)/\rho(0) >4$ in Si MOS systems in
the valley polarized situation.

\begin{figure}
\epsfysize=2.4in
\centerline{\epsffile{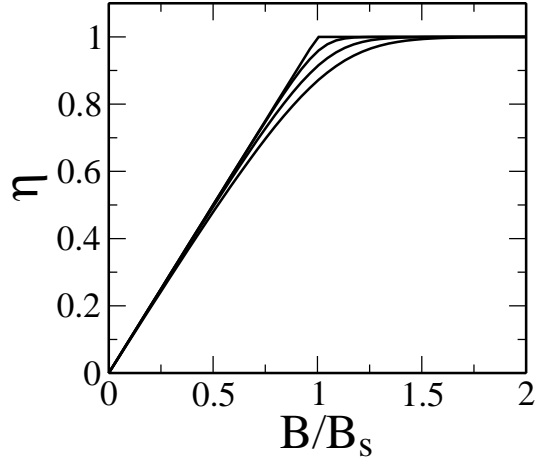}}
\caption{
Spin-polarization $\eta = (n_+ -n_-)/(n_+ + n_-)$ 
as a function of in-plane magnetic field for $n=1.5 \times
10^{11}cm^{-2}$ and for different temperatures $T=0$, 1.0, 2.0, 3.0K.
$B_s$ is the T=0K full spin-polarization field.
Note that for $T \neq 0$, full spin-polarization requires $B \approx 2
B_s$. }
\end{figure}

There is, in fact, a qualitative physical
explanation \cite{shashkin1}
for the observed temperature independence of the high-field
magnetoresistance in Si MOSFETs, which is generic and universal in
nature and does not invoke the ad hoc explanation of the field
-induced lifting of the silicon valley degeneracy we propose above as
a possibility. This explanation is, however, based on a categorical
repudiation of the existing experimental claims \cite{tsui} that
the constancy (i.e. temperature independence) of $\rho(T;B>B_s)$ as a
function of temperature at high parallel fields does not in any way
indicate a fundamental suppression of ``metallicity'' (i.e. the
metallic temperature dependence) in the {\it fully} spin-polarized Si
MOSFETs, as has been repeatedly emphasized by several experimental
groups in the past \cite{tsui,shashkin,review1}. 
Instead, the rather generic 
explanation, recently suggested in ref. \onlinecite{shashkin1}, is that
the observed (essentially) complete suppression 
of the metallic temperature dependence occurs at fields
{\it below} the full spin polarization field (at finite temperature)
where the 2D system is still partially spin-polarized. 
In Fig. 3 we show the calculated spin-polarization as a function of
the in-plane magnetic field for different temperatures. 
As the temperature increases the net spin polarization at $B=B_s$
decreases due to the thermal excitation.
The suppression of ``metallicity'' in this simple explanation arises
from two competing temperature-induced mechanisms in the partially
spin-polarized 2D system at parallel fields just below complete
saturation. The two mechanisms counter-balance each other because one
increases screening and the other decreases screening. Below (but
close to) the full spin-polarization field $B \le B_s$, increasing
temperature {\it reduces} screening through the direct thermal
broadening, but {\it enhances} screening by thermally exciting
reversed-spin quasiparticles (i.e. by thermally reducing the net spin
polarization). Actually, these two competing temperature-induced
mechanisms are always present in the 2D carrier system at any finite
parallel field: increasing temperature induces two competing trends
in screening -- increased thermal broadening reduces screening and
thermal excitation enhances screening by reducing the net spin
polarization. Our theory and the numerical results, of course, include
these two mechanisms. 

\begin{figure}
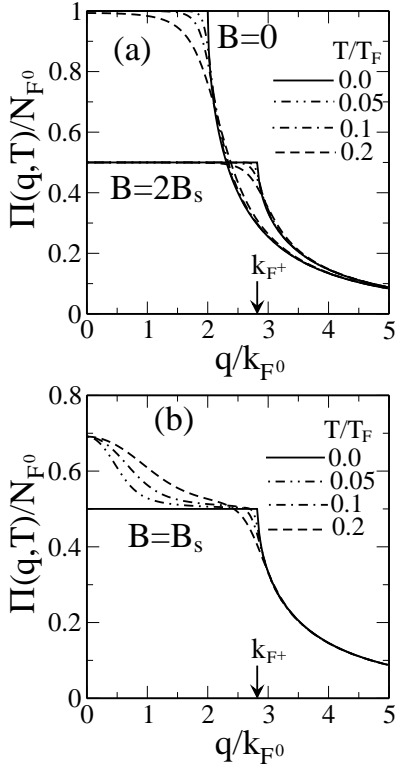

\epsfysize=2.in
\centerline{\epsffile{fig4a.eps}}
\centerline{\epsffile{fig4b.eps}}
\caption{The calculated polarizabilities for different
in-plane magnetic fields (a) for $B=0$ and $B=2B_s$, and (b) for
$B=B_s$. 
Here $N_{F^0}=2N_F=g_v m/2\pi$ is the density of states for
unpolarized system, and $k_{F^+} = \sqrt{2}k_{F^0}$.
}
\end{figure}

In Fig. 4 we show the calculated polarizabilities, $\Pi(q,T)/N_{F^0}$
($N_{F^0}=2N_F=g_v m/2\pi$),
for different in-plane magnetic fields; (a) for $B=0$ and $B=2B_s$,
and (b) for $B=B_s$. 
At $B=0$ (i.e., the system is unpolarized and $n_+ = n_-$),
the decrease of the screening function (i.e. $\Pi$) with increasing
temperature at
$q=2k_{F^0}$ ($k_{F^0}=\sqrt{2\pi n/g_v}$) gives rise to increasing
resistivity as the 
temperature increases. Most of the temperature dependence of the
resistivity at low temperature regime comes from the suppression of 
screening at $q=2k_{F^0}$. The reduction of screening near
$q=2k_{F^0}$ determines the strength of metallicity at $B=0$. 
At $B=2B_s$ (i.e.,
the system is completely polarized and $n_+ = n$ at $T=0$), we find
a very similar behavior of the screening function, but the
rate of the change of screening at $q=2k_{F^+}$
($k_{F^+}=\sqrt{4\pi n/g_v}$) is
reduced by about a factor of four 
compared to the $B=0$ case in the same
temperature range.  Thus, we expect that the temperature dependence of
$\rho(T)$ at $B=0$ is roughly a factor of 4 stronger than that at
$B=2B_s$. At $B=2B_s$ the population of the
minority band due 
to the thermal excitation is negligible in the given temperature range.
In Fig. 4(b) we show the polarizability function at $B=B_s$. 
As the temperature
increases the screening function decreases at $q=2k_{F^+}$ for $B=B_s$,
but due to the occupation of the minority band we also 
find an enhancement
of the screening function near $q=2k_{F^-}$ ($k_{F^-}=\sqrt{4\pi
  n_-/g_v} < k_{F^0}$). Thus, the enhancement 
of the screening function near
$2qk_{F^-}$ with increasing of the minority spin carrier density
and the reduction of
screening near $q=2k_{F^+}$ by thermal broadening 
give rise to weak temperature dependence in $\rho$ around $B \sim
B_s$. 

\begin{figure}
\epsfysize=1.55in
\centerline{\epsffile{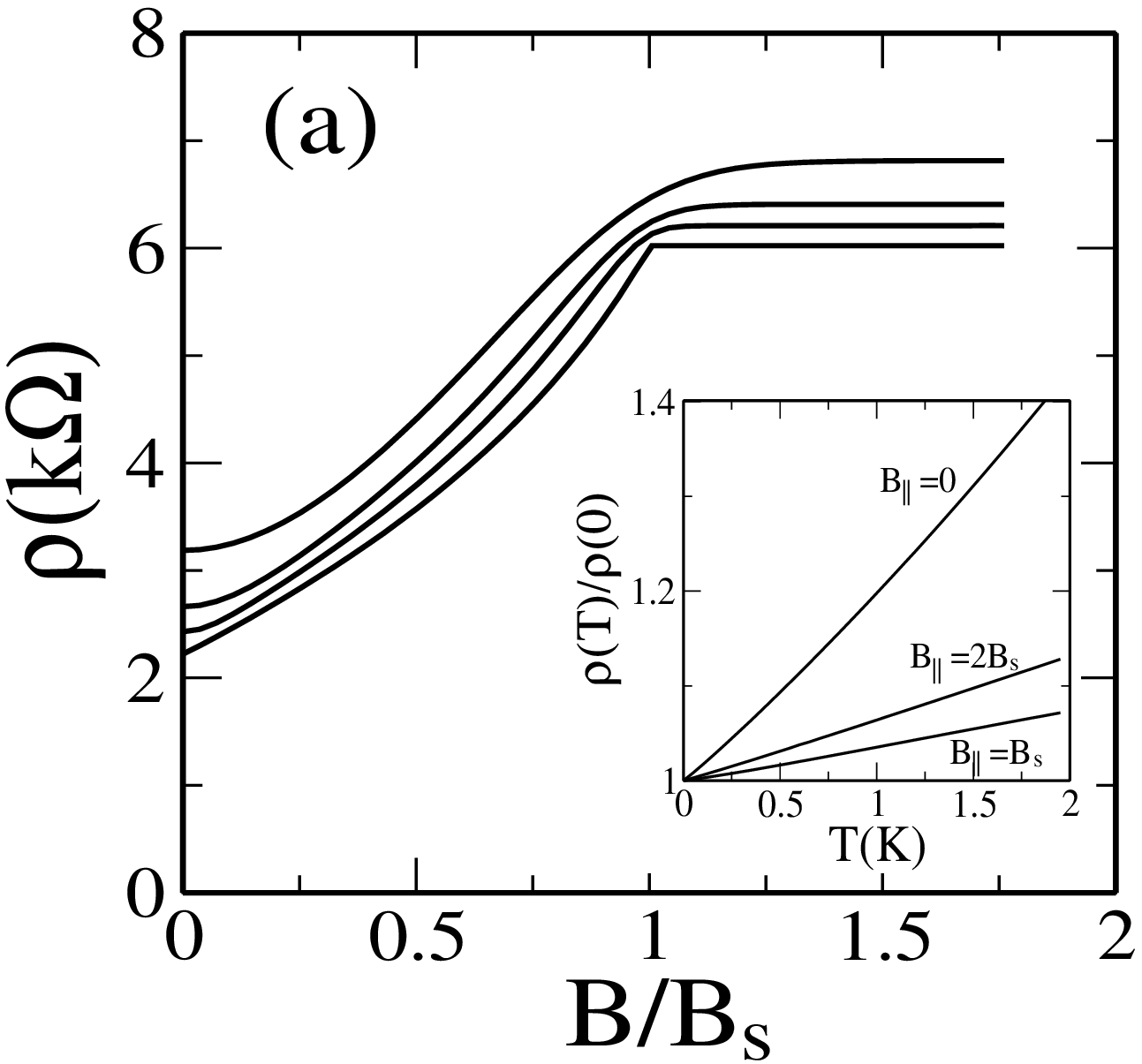}}
\centerline{\epsffile{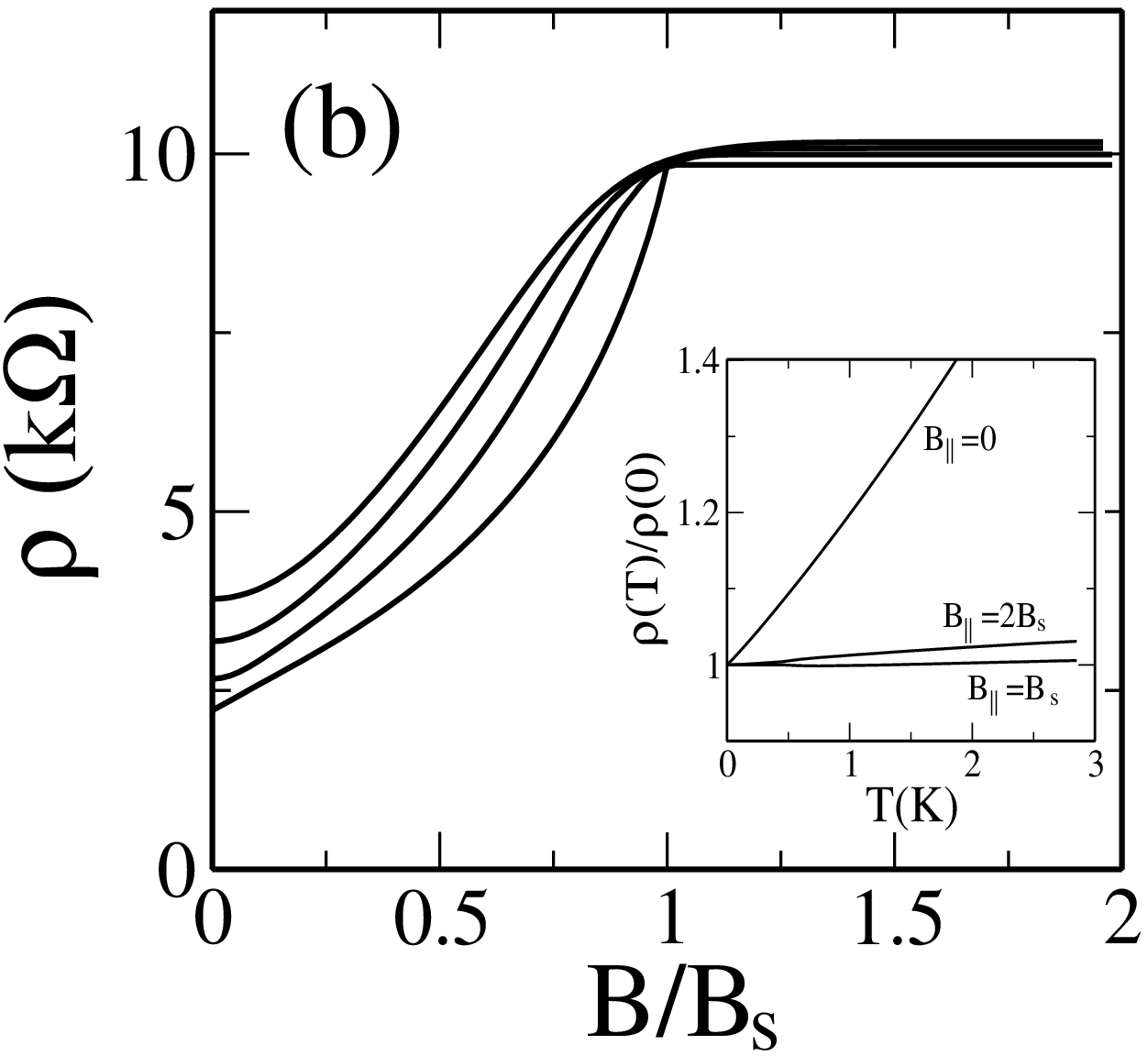}}
\caption{
Calculated magnetoresistance as a function of in-plane magnetic field
for $n=1.5 \times 10^{11}cm^{-2}$.
Fig. (a) shows the results where the valley degeneracy is not affected
by the parallel field, 
i.e. $g_v = 2$ for all magnetic fields, but in (b) the valley
degeneracy varies with $B_{\|}$ in the same manner as the spin
degeneracy. 
In (a) the lines correspond to  the results for $T=0.0$,
0.5. 1.0 2.0K (from bottom to top).  In (b) we use $T=0.0$, 1.0, 2.0,
3.0K (from bottom to top).
Insets show the normalized resistivity as a function of temperature at
fixed magnetic fields $B_{\|}=0$, $B_s$, and $2B_s$. 
}
\end{figure}

It is, in principle, possible for these two competing
mechanisms to almost completely cancel each other, close to $B \approx
B_s$.
In Fig. 5 we show our numerical results
for both situations of the valley degeneracy being affected (and
unaffected) by the parallel field. Obviously the temperature
independence of $\rho(T)$ around $B \sim B_s$ is more prominent in the
situation where the valley degeneracy (as well as the spin degeneracy)
is lifted by the external magnetic field. We emphasize that these
competing temperature induced trends in screening in the presence of a
finite spin-polarizing parallel magnetic field (i.e. the direct
suppression of screening by increasing temperature and the indirect
enhancement of screening by the temperature induced decrease of
effective spin-polarization) is a generic phenomenon, and is therefore
present in {\it all} 2D carrier systems. In particular, these
competing mechanisms could be the underlying cause for the occurrence
of the so-called parallel field-induced 2D metal-insulator transition,
which has been widely discussed in the context of hole transport in
high-mobility 2D p-GaAs structures. It may, therefore, be possible
that the peculiar temperature independence of 2D resistivity around $B
\sim B_s$ observed in n-Si MOSFETs arises from both the parallel field
induced lifting of valley degeneracy as well as the competing
mechanisms discussed above (since valley degeneracy is the only
physical mechanism that distinguishes the Si MOS system from all the
other 2D semiconductor systems).

It is important to emphasize that our finding (see Fig. 5) of the
temperature dependence of $\rho(T,B)$ being the weakest below the
full-spin-polarization field (at finite temperature), i.e. for $B
\approx B_s$, and the strongest for $B=0$ with the fully
spin-polarized system ($B \gg B_s$) having an intermediate behavior,
as observed experimentally in ref. \onlinecite{shashkin1},
is valid independent of whether the Si valley degeneracy is lifted by
the applied magnetic field or not. This is a generic result which is
always true in all 2D systems independent of the number of valleys:
The temperature dependence of $\rho(T,B)$ is always the weakest for $B
\approx B_s$ due to the competition between the thermal excitation
between spin up/down bands and the thermal suppression of
screening. This is obvious from Fig. 5 where the temperature
dependence of $\rho(T,B)$ is the weakest around $B \sim B_s$, both in
Fig. 5(a) and in Fig. 5(b) with the assumed valley degeneracy being 2
and 1 respectively. As we have emphasized above, the additional
assumption of a magnetic field induced lifting 
of the valley degeneracy makes
$\rho(T)$ almost a constant, completely independent of temperature, in
quantitative agreement with the data of Tsui {\it et al.}
\onlinecite{tsui}. Whether this actually happens or not can only be
decided experimentally \cite{pudalov_g}. We should, however, emphasize
that even if the valley splitting $\Delta_v$ is small (e.g. $\Delta_v
< 2E_F$ so that the valley degeneracy is not completely lifted), a
partial lifting of valley degeneracy ($\Delta_v \ne 0$) will certainly
contribute to the physics being proposed here. For a partial lifting
of the valley degeneracy, our calculated temperature dependence of
$\rho(T)$ will be intermediate between that shown in Figs. 1(b) and
1(c) or Figs. 5(a) and 5(b) as the case may be. In particular, a
finite valley splitting will lead to the same competing thermal trends
in $\rho(T)$ as discussed above and by Shashkin {\it et
  al}. \cite{shashkin1} for the spin splitting, producing an
additional suppression of $\rho(T)$ bringing our theoretical results
to better quantitative agreement to the Tsui {\it at
  al}. \onlinecite{tsui} data.

\section{discussion}

We have shown in this paper that the enigmatic temperature dependence
of the parallel field magnetoresistance, $\rho(T,B)$, in 2D Si MOSFETs
can be qualitatively and semiquantitatively well explained by the
screening theory, assuming charged impurity scattering to be the
dominant low-temperature resistive scattering mechanism. Our screening
theory based explanation, in fact, becomes quantitatively accurate if
we {\it assume} that the valley degeneracy is lifted by the external
magnetic field in the same manner as the spin degeneracy. The fully
spin- and valley-polarized Si MOS system at high field has a factor of
4 lower density of states than the corresponding zero-field spin- and
valley-unpolarized system, leading consequently to substantially weaker
screening with the corresponding strong suppression in the temperature
dependence of the resistivity. 
We show that, independent of the valley degeneracy question, 
the weakest temperature
dependence in the resistivity happens {\it not} at $B \gg B_s$ (where
the carriers are essentially completely spin-polarized), but at $B \le
B_s$, just below the $T=0$
full spin-polarization field, where the system is
almost, but not quite, fully spin-polarized. Thus, the metallicity
(i.e. the magnitude of $d\rho/dT$) is the strongest for $B=0$ and the
weakest for $B \le B_s$ with the fully spin-polarized ($B \gg B_s$)
situation being intermediate. This specific prediction of the screening
theory is in good agreement with experimental observations both in
n-Si MOSFETs \cite{shashkin1} and 
in n-Si/SiGe \cite{sige} 2D systems. The key feature of our
theory is a realistic and quantitatively accurate description of 2D
screening taking into account finite temperature and finite
spin-polarization effects on an equal (and non-perturbative) footing.
The important ingredient of physics leading to the strong suppression
of metallic temperature dependence of $\rho(T,B\approx B_s)$ in our
theory is a competition between two physical mechanisms: Thermal
excitation of reversed spin quasiparticles leading to enhanced
screening with increasing temperature for $B \le B_s$ and the direct
thermal suppression of screening by the majority spin carriers. We
note that this competition would always lead to a strong suppression
in the temperature dependence of $\rho(T)$ around $B \le B_s$,
independent of the member of valleys involved in the 2D system.  
It is also important to emphasize that for $B > B_s$ (and at not too
low carrier densities) our theory predicts a {\it negative}
magnetoresistance with monotonically decreasing $\rho(B)$ with
increasing $B$ due to the Coulomb matrix element effect arising from
an increasing effective $k_F(B)$ in the system. This also leads to the
suppression of metallic temperature dependence at high fields.

Our finding of the qualitative 
agreement between the screening theory and the
experimentally observed $\rho(T;B)$ in Si MOSFETs is of considerable
importance. Much has been made in the literature of the
claimed non-metallic (i.e. temperature independence or even
insulating) behavior of 2D magnetoresistatce in Si MOSFETs at high ($B
\ge B_s$) parallel fields. In particular, interaction-based theories
\cite{ZNA} of 2D metallicity have been claimed \cite{review1} to
provide the definitive explanation for 2D metallicity in Si MOSFETs
because the interaction theory predicts the fully spin and
valley-polarized 2D 
system to have an insulating (with $\rho \propto 1/T$) temperature
dependence, and it has been categorically asserted that the screening
theory cannot possibly be qualitatively correct \cite{comment} since
it predicts a (weak) metallic behavior in the fully spin-polarized
ballistic regime.
We have shown in this work that the screening theory is, in fact, in
good qualitative (and even quantitative, if we allow the possibility
of valley degeneracy lifting) agreement with the most recent Si MOS
experimental data \cite{tsui,shashkin1} in the presence of a parallel
magnetic field, and in particular, consistent with our theoretical
predictions, the strongest suppression of the metallic temperature
dependence of resistivity happens for $B \le B_s$ where the 2D system
is almost, but not quite, fully spin polarized. Thus, $\rho(T,B)$
manifests the strongest metallicity at $B=0$ (when the system is
unpolarized) the weakest at $B \approx B_s$ (when the system is almost
polarized), and intermediate at $B \gg B_s$ (when the system is fully
polarized) both in the screening theory and in the experiment. Exactly
the same behavior has also been seen recently by two experimental
groups \cite{sige} in 2D n-Si/SiGe electron systems which we have
theoretically explained \cite{sige_hwang} recently using the screening
theory. Thus, the parallel field induced suppression of metallic
temperature dependence in the ballistic regime is now theoretically
understood (at least qualitatively) for
both n-Si MOS and n-Si/SiGe 2D electron systems, and the early
experimental discrepancies between these two systems have now been
resolved \cite{shashkin1} with both systems agreeing reasonably with
the screening theory predictions in the ballistic regime.

There are various proposed non-Fermi liquid scenarios for the 2D
metallicity which depend crucially on important distinctions
between the spin polarized ($B > B_s$) and the unpolarized ($B=0$)
transport regimes. At least one of these non-Fermi liquid scenarios,
based on a speculative coexistence between 2D Wigner crystal and
electron liquid phases \cite{spivak} is  
apparently invalidated by the
observed weak metallicity in the spin-polarized Si MOSFETs, since the
theory \cite{spivak} specifically predicts absolutely no temperature
dependence in $\rho(T; B\gg B_s)$ in the spin-polarized phase in
direct contradiction with the recent experimental results.
\cite{sige,shashkin1}

We briefly discuss the role of interaction in our screening
theory, touching upon the closely related (and important)
question of why the Boltzmann-RPA screening theory seems to provide a
good description of the 2D transport properties at low carrier
densities where the dimensionless interaction parameter $r_s$
($\propto n^{-1/2}$ in 2D), defined as the ratio of the average
Coulomb energy to the Fermi energy at $T=0$, is larger than one (in
Si-based 2D systems of interest in this paper, $r_s \sim 6$ in the
experimentally relevant carrier density range of $n \sim
10^{11}cm^{-2}$). We believe that one possible reason for the success
of the RPA screening theory is that the dominant contribution to the
2D resistivity arises from charged impurity scattering, and
RPA-Boltzmann theory does an excellent job of regularizing the
associated Coulomb disorder through the self-consistent screening
model. The 2D transport problem is, in fact, semiclassical since the
temperature, expressed in units of the Fermi temperature, is {\it not}
necessarily small (often $T/T_F \sim O(1)$), again making finite
temperature RPA an excellent approximation. It is important to
emphasize that at a fixed finite temperature $T$, decreasing 2D
density increases both $r_s$ and $T/T_F$ (since $T_F \propto n$), and
the finite-temperature interaction parameter, e.g. $r_s/(T/T_F) \sim
\sqrt{n}$, actually decreases with decreasing density! Therefore, it
is not obvious at all that finite temperature 2D transport behavior
increasingly becomes strong-coupling like as electron density is
decreased at a fixed temperature since the low-density ($r_s
\rightarrow \infty$) limit is also at the same time the classical
infinite temperature ($T/T_F \rightarrow \infty$) limit. Under these
circumstances, the semiclassical RPA-Boltzmann finite temperature
theory may, in fact, become increasingly more valid as the carrier
density is reduced, particularly since the carrier temperature can
often not be reduced below 100 mK due to electron heating
effects. This finite-temperature aspect of the low-density 2D
transport problem, which has not been adequately emphasized in the
literature, may very  well be playing an important role in making RPA
a particularly good approximation in the 2D MIT phenomena. 

We note that RPA-based many-body theories also seem to describe
reasonably well \cite{hwang_pl,R2} the experimental behaviors of
finite-temperature, dilute 2D carrier systems with respect to
collective mode dispersion \cite{R3} and inter-layer drag measurements
\cite{R4}. This
finite-temperature (actually {\it high-temperature}) aspect of this
problem most likely
also invalidates any possible relevance of  Wigner
crystallization to the physics of 2D MIT phenomena -- a Wigner crystal
would thermally melt \cite{hwang_pl,R5} at the ``relatively''
high temperatures (i.e. $T/T_F \sim O(1)$) at which the low density 2D
transport experiments are typically carried out.
It needs to be emphasized that the asymptotic low-temperature behavior
of $\rho(T)$ is essentially never observed in 2D transport experiments
since the experimental $\rho(T)$ always saturates at the lowest
measurement temperatures (often for $T \le 100$mK) indicating that the
electrons may not be cooling down to the $T/T_F \ll 1$ regime. In this
intermediate to high temperature regime our screening theory, which is
non-perturbative in temperature, may very well be the qualitatively
correct approximate theory. Understanding the true asymptotic
temperature dependence of $\rho(T)$ in the $T/T_F \ll 1$ regime, where
interaction effects must be important in a low-density strong-coupling
electron system, remains an important open experimental and
theoretical challenge.

We can speculate on the possibility of 
incorporating interaction effects into our
screening theory of 2D transport. The 2D resistivity, $\rho(n,T,B)$,
depends on density, temperature, and the applied magnetic field, which
may be expressed in the dimensionless units to write
\begin{equation}
\rho(n,T,B) = \rho(q_{TF}/2k_F, T/T_F, \Delta_s/T_F),
\end{equation}
where $\Delta_s = g \mu_B B$ is the spin splitting induced by the
external magnetic field. We note that $q_{TF}/2k_F \propto m$; $T/T_F
\propto m$; and $\Delta_s/T_F \propto gm \propto \chi$ where $m$, $g$,
and $\chi$ are respectively the 2D effective mass, Lande $g$-factor,
and spin susceptibility. In the RPA theory, interaction effects are
ignored in the effective mass and the $g$-factor, and therefore, the
noninteracting effective mass and the susceptibility are used in
calculating the resistivity. In the spirit of the Landau Fermi liquid
theory, we could crudely incorporate interaction effects in the theory
by using the quasiparticle effective mass ($m^*$) and the quasiparticle
spin susceptibility ($\chi^*$) or equivalently the quasiparticle
$g$-factor ($g^*$) in the effective RPA calculation. A rigorous
justification for such a Fermi liquid renormalization of the effective
mass and $g$-factor in the RPA-Boltzmann transport theory is
unavailable, and indeed all our numerical results utilize the
non-interacting band mass and $g$-factor in the Boltzmann-RPA
calculations, but we speculate that such a physically motivated
approximation (i.e. $m^*$ and $g^*$ replacing $m$ and $g$) to
incorporate interaction effects may be quite reasonable since the
observed or measured quantities are actually $m^*$ and $g^*$ (and
$\chi^* = m^*g^*$) and {\it not} the bare quantities ($m$ and $g$). It
is interesting to point out that such an {\it ad hoc} approximation
scheme incorporating interaction effects (by using $m^*$ and $g^*$ in
the Boltzmann-RPA transport theory) does considerably improve the
quantitative agreement between theory and experiment. In particular,
the interacting susceptibility $\chi^*$ is renormalized \cite{zhang} 
by a factor of
3 (i.e. $\chi^*/\chi \sim 3$) at $n \sim 10^{11}cm^{-2}$ in Si MOSFETs
leading to the theoretical spin-polarization saturation magnetic field
$B_s$ ($\sim 3-4$T) being in good agreement with the experimental
data. Similarly, the use of the quasiparticle effective mass $m^* \sim
3m$ at low densities considerably increases the effective values of
$q^*_{TF}/2k_F$ and $T/T_F^*$ (where ``starred'' quantities use $m^*$
rather than $m$), again bringing theory and experiment in good
quantitative agreement. Whether such an {\it ad hoc} ``improvement'' of
the theory using quasiparticle rather than bare Fermi liquid
parameters can be rigorously theoretically justified or not remains an
important open question for the future. We note, however, that the
screening theory obtains  semi-quantitative and qualitative
agreement with the existing 2D transport data even without any such
``renormalization''.

Finally, we note that there still seems to be some quantitative
difference in the observed experimental temperature dependence of
$\rho(T;B)$ between Si MOS \cite{tsui,shashkin} and Si/SiGe
\cite{sige} 2D electron systems. 
For example, in the Si MOS system $\rho(B)$ is enhanced approximately
by a factor of three at low temperatures as $B$ increases from zero to
$B_s$ whereas in the Si/SiGe system \cite{sige} the enhacement factor
is only almost 1.8. 
This difference arises (at least
partially) from the difference in the charged impurity distribution in
the two systems --- in the Si/SiGe 2D electron system the dominant
scattering is from {\it remote} dopants and background impurities
\cite{sige_hwang} whereas in Si MOS system the scattering is mostly by
interface charged impurities and interface roughness. In addition, the
finite magnetic field behavior in these two Si-based 2D systems may
also differ by virtue of the valley degeneracy being lifted in the Si
MOS system, but {\it not} in the Si/SiGe system. Such a non-universal
lifting of the valley degeneracy is certainly possible since it is
well-known that the Si valley degeneracy near an interface depends
critically on the microscopic details of the interface, and it is
entirely possible for the valley degeneracy to be lifted in the Si MOS
system, but {\it not} to be lifted in the Si/SiGe system since the
latter system has a much better atomically smooth epitaxial
interface. In fact, the non-universal lifting of valley degeneracy may
apply even to different Si MOS 2D systems, and could provide one
possible underlying reason for the observed difference 
\cite{tsui,shashkin} in the temperature dependence of $\rho(T;B\ge
B_s)$ in Si MOS data from different groups.
Whether such a non-universal parallel field-induced
electron valley degeneracy lifting actually happens in reality can
only be decided experimentally; we are only
suggesting here the theoretical possibility based on our analysis of
the experimental transport properties of $\rho(T;B)$.

\section{conclusion}

In conclusion, we consider theoretically the parallel magnetic field
induced suppression of the screening of long range bare Coulomb
disorder in Si MOSFETs, showing that the experimentally observed
strong suppression of 2D metallicity in the temperature dependence of
the magnetoresistance 
can be qualitatively and semiquantitatively understood as arising
from spin (and perhaps even valley) polarization induced reduction in
carrier screening, leading to stronger parallel field-dependent
effective disorder in the system.
The competing mechanism of direct thermal reduction of screening and
indirect enhancement of screening through the thermal suppression of
spin polarization may also be playing an important quantitative role
in the temperature dependence of $\rho(T;B\sim B_s)$. Our theory, 
as presented in the current work and in our recent work
\cite{sige_hwang} on the 2D Si/SiGe electron system 
along with the recent experimental work \cite{sige,shashkin1},
may resolve the earlier
discussed qualitative disagreement between Si MOS and Si/SiGe 2D
systems, establishing the same qualitative behavior in all 2D Si
systems. More work is still needed to precisely understand interaction
effects on 2D transport by going beyond the physically motivated
screening theory of our work.

This work is supported by ONR, NSF, and LPS.

\end{document}